\documentclass[aps,prx,reprint,twocolumn,floatfix,citeautoscript,longbibliography,superscriptaddress]{revtex4}
\usepackage{graphicx}
\usepackage{bm,amsmath,amssymb,mathrsfs,dcolumn}
\usepackage[colorlinks, linkcolor=blue, citecolor=blue, urlcolor=blue, linktoc=page, bookmarks=false, pdfstartview={FitH}, pdfborder={0 0 0.0 [3 3]}]{hyperref} 
\usepackage[usenames]{color}
\hypersetup{pdfborder=0 0 0,colorlinks=true,citecolor=blue,linkcolor=blue}
\usepackage[final]{pdfpages}
\usepackage{epstopdf}
\usepackage{subfigure}
\usepackage{units}
\usepackage{url}
\usepackage{float}
\usepackage{natbib}
\usepackage{graphicx}

\begin{document}

\title{Giant magnetoresistance in antiferromagnetic Mn$_2$Au-based tunnel junction}
\author{Xing-Tao Jia}
\altaffiliation{Corresponding author, E-mail:jiaxingtao@hpu.edu.cn}
\author{Xiao-Lin Cai}

\affiliation{School of Physics and Electronic Information Engineering, Henan Polytechnic
University, Jiaozuo 454000, China}

\author{Yu Jia}
\affiliation{International Laboratory for Quantum Functional Materials of Henan, and School of Physics, Zhengzhou University, Zhengzhou 450001, China.}
\affiliation{Key Laboratory for Special Functional Materials of Ministry of Education, School of Materials Science and Engineering, Henan University, Kaifeng 475004, China.}

\date{\today }

\begin{abstract}
Recent studies on the electrical switching of tetragonal antiferromagnet (AFM) via N{\'e}el spin-orbit torque have paved the way for the economic use of antiferromagnetic materials. The most difficult obstacle that presently limits the application of antiferromagnetic materials in spintronics, especially in memory storage applications, could be the small and fragile magnetoresistance (MR) in the AFM-based nanostructure. In this study, we investigated the spin transports in Mn$_2$Au-based tunnel junctions based onthe first-principle scattering theory. Giant MRs more than $1000\%$ are predicted in some Fe/MgO/Ag/Mn$_2$Au/Ta junctions that are about the same order as that in an MgO-based ferromagnetic tunnel junction with same barrier thickness. The interplay of the spin filtering effect, the quantum well resonant states, and the interfacial resonant states could be responsible for the unusual giant and robust MRs observed in these Mn$_2$Au-based junctions.
\\


pacs: 72.25.¨Cb, 75.47.De, 75.50.Ee, 85.75.Mm
\end{abstract}

\maketitle

\section{Introduction}

One of the most thrilling challenges that currently limit the application of antiferromagnetic spintronics is the lack of large and robust magnetoresistance (MR) effect comparable to that in ferromagnetic tunnel junctions (F-MTJs).\cite{gomonay2014spintronics,jungwirth2016antiferromagnetic,baltz2018antiferromagnetic,vzelezny2018spin} The anisotropic MR (AMR) in the antiferromagnet (AFM) has been adequately
studied; it exhibits a relativistic effect and  is about the same order as that in the ferromagnet (FM). The tunneling effect can considerably enhance the AMR effect.\cite{park2011spin,wang2012room} AMR larger than $100\%$ at $4~\unit{K}$ was reported in an NiFe/IrMn/MgO/Pt junction that  was reduced by a few percent at $100~\unit{K}$.\cite{park2011spin} MRs in the AFM-based spin valves have been theoretically studied for a long time.\cite{nunez2006theory,xu2008spin,saidaoui2014spin,jia2017structure,PhysRevApplied.12.044036} MRs in the AFM-based spin valves  are smaller in magnitude than that in the MgO-based F-MTJ.\cite{butler2001,mathon2001theory,ikeda2010perpendicular} For high-density memory applications, the structure could be resistive, electric current-driven, and it should have MR not less than $100\%$.\cite{parkin2003magnetically,chappert2010emergence} Unfortunately, no AFM-based structure can satisfy all these demands at the same time at the moment.

Several methods are used to control the N{\'e}el order in AFMs.\cite%
{0957-4484-29-11-112001,gomonay2014spintronics,jungwirth2016antiferromagnetic,baltz2018antiferromagnetic,vzelezny2018spin}
Among these methods, the use of electric current is more favorable for compatibility
with the present state-of-the-art semiconductor technology. The spin dynamics in
the AFM can be depicted by coupled Landau-Lifshitz-Gilbert equations
with a working frequency of up to $\unit{THz}$\cite{jungwirth2016antiferromagnetic,baltz2018antiferromagnetic,vzelezny2018spin} that is around
three orders faster than that of the FM. Owing to three order reduction in the spin dynamics time, the switching of the N{\'e}el order of the AFM should show more power-saving ability than that of the magnetization of the FM. Furthermore, the antiferromagnetic materials also have the advantage of robust stability in magnetic fields, no stray fields, and large magneto-transport effects.

Here, we pay attention to the MRs in the Mn$_2$Au-based antiferromagnetic tunnel junctions (AF-MTJs), where the N{\'e}el order of Mn$_2$Au is set free. Bulky Mn$_2$Au has a tetragonal structure\cite{barthem2013revealing}
 of space group $I4/mmm$ with $a=3.328~\mathring{A}$ and $c=8.539~\mathring{A}$.  The magnetic sublattices of tetragonal Mn$_2$Au are collinear with broken inversion symmetry,\cite{Wadley587} which makes it possible to switch the N{\'e}el order via N{\'e}el spin-orbit torque (NSOT).\cite{vzelezny2014relativistic,Wadley587,bodnar2018writing,meinert2018electrical,PhysRevLett.120.237201,chen2019electric} Moreover, the spin-orbit torque can be used to switch the  N{\'e}el order as well.\cite{zhou2019fieldlike,PhysRevApplied.9.054028} Metallicity in synergy with high N{\'e}el
temperature above $1000~\unit{K}$\cite{barthem2013revealing}
makes Mn$_2$Au a promising candidate for antiferromagnetic spintronic
applications.\cite{barthem2013revealing,wu2016anomalous,bodnar2018writing,meinert2018electrical,PhysRevLett.120.237201}
Also, large AMR of $\sim6\%$ was reported in thin Mn$_2$Au films.\cite{wu2016anomalous,bodnar2018writing}
The huge tunneling MR (TMR) of over $1000\%$ in the MgO-based F-MTJ is related to the perfect spin filtering effect,\cite{butler2001,mathon2001theory} which is responsible for the large MR $\sim100\%$
in the FeMn-based AF-MTJs.\cite{jia2017structure} When a Mn$_2$Au film is
contacted with an MgO film to form an interface, the L-type magnetic structure with staggered spins along the direction of the current flow would enhance the MR.\cite{nunez2006theory,jia2017structure} In this study, we report first-principles investigations of the spin transports of the Fe/MgO/Ag/Mn$_2$Au/Ta AF-MTJs. Giant MRs of about the same order as that in the MgO-based F-MTJs were found at some junctions.

\begin{figure}[tbp]
\centering
\includegraphics[width=8.6cm]{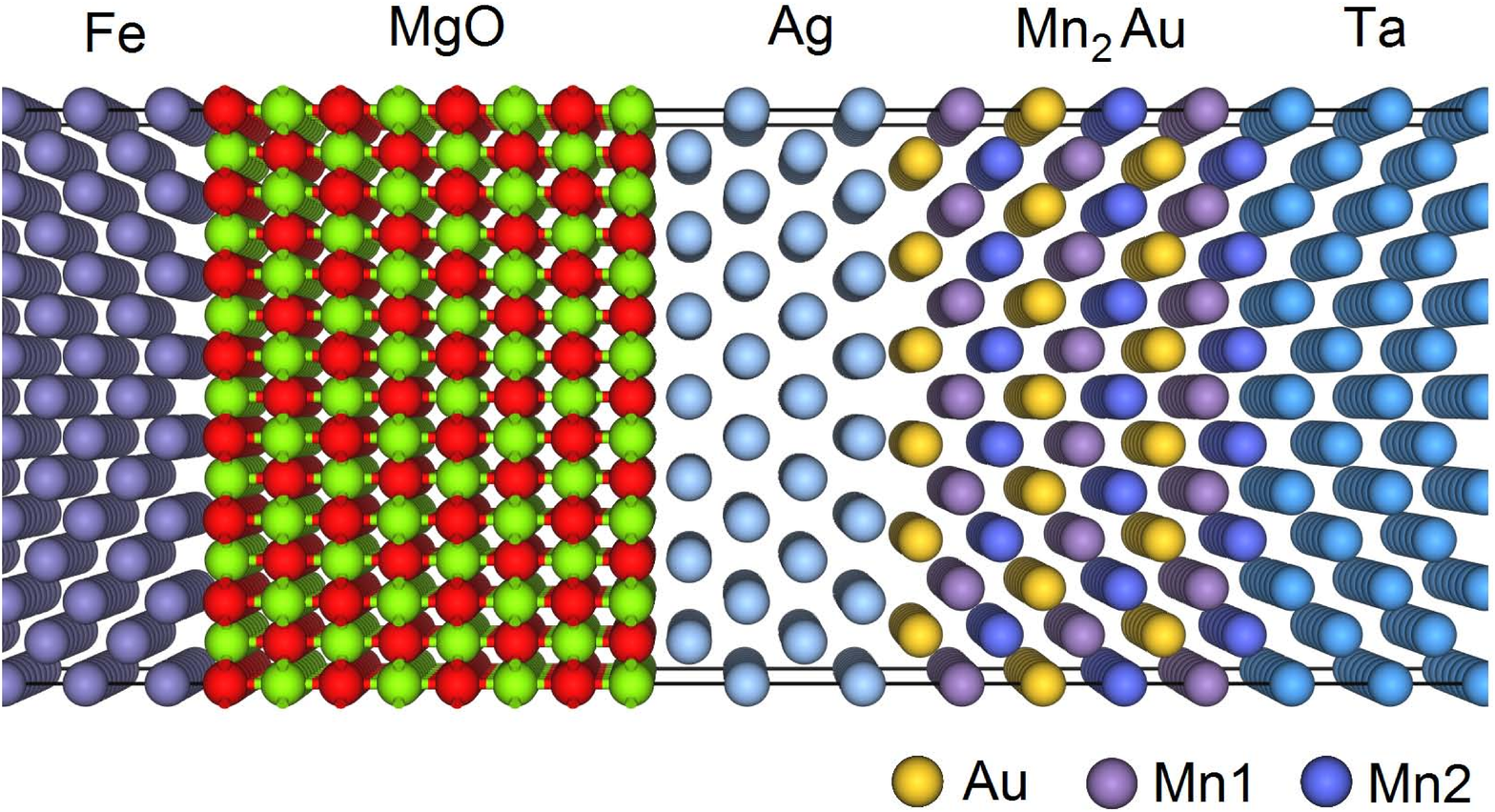}
\caption{Schematic Fe/MgO/Ag/Mn$_2$Au/Ta(001) AF-MTJ used in the study. Therein, Mn1 and Mn2 represent Mn atoms with positive and negative magnetic moments, respectively. The left-side magnetization was fixed, and the right-side N{\'e}el order sets free, which follows the magnetization direction of the Mn atoms that were in contact  with the Ag layer.}
\label{fig1}
\end{figure}

\section{Methods}

A two-terminal structure with MgO/Ag/Mn$_2$Au sandwiched between the left semi-infinite Fe lead and the right semi-infinite Ta lead was used to study the MR effect as shown in Fig. \ref{fig1}. The spin transport calculations were carried out via a first-principle
Wave-Function-Matching method\cite{wang2008first} with the potentials
obtained via self-consistent calculations performed using the tight-binding
linear muffin-tin orbital (TB-LMTO) surface Green's function method with
a coherent potential approximation to deal with the disorder or impurities
.\cite{turek1997electronic} The modified BJ potential\cite{tran2009accurate}
was used for Mn$_2$Au during the self-consistent calculations, and this showed that the moments of the Mn atoms was $\pm3.96~\mu_B$ and the moment of Au
atoms was almost zero $\mu_B$. This is consistent with recent experiment\cite{barthem2013revealing} and it aids improvements than the LDA potential, and this showed that the moments of Mn atoms are $\pm3.62~\mu_B$. During the calculations, a $7\times 7$ Fe/MgO/Ag lateral supercell was used to match a $6\times 6$ Mn$_2$Au/Ta lateral supercell, as shown in Fig. \ref{fig1}. The Ag/Mn$_2$Au interface was relaxed to minimize the volume overlap of the interfacial atoms. A 40$\times $40 \textit{k}-mesh was used to sample the two-dimensional Brillouin zone (2d BZ) to ensure good numerical convergence. For the junction with imperfections, such as the interfacial Oxygen vacancy (OV), interfacial alloy, and thermal lattice disorders, over 20 configurations were used to obtain average convergence.

\section{Results and Discussions}

\begin{table}[tbp]
\caption{The conductance $G$ in units of $10^{12}~\Omega^{-1}\unit{m}^{-2}$ in the Fe/MgO(5)/Ag(4)/Mn$_2$Au(12)/Ta junctions with Au-term structure with respect to the minimal distance $d$ between the Ag atoms and the Au(Mn) atoms in the Mn$_2$Au layer. Four interfacial atomic layers (two Ag layers and two Mn$_2$Au layers) are relaxed in scheme II, III, IV, and V.}
\label{tab2}%
\begin{tabular*}{8.6cm}{@{\extracolsep{\fill}}lcccccc}
\hline\hline
Scheme &$d~(\AA)$& $G$(PC)& $G$(APC) &  MR$~(\%)$ & Overlap$^1$ & Overlap$^2$ \\ \hline
I$^3$ &1.73 &0.659  &0.0528  & 1150  & 0.212& 0.06948 \\
II & 1.91& 0.701 &  0.0497 &  1310  & 0.217& 0.06964 \\
III & 2.08&0.636 &0.0566  & 1060  & 0.214& 0.06946 \\
IV  &2.34 &0.305  &0.0291 & 950  & 0.222&0.07004  \\
V&2.61 &0.364  &0.0268  & 1260  & 0.226&0.07330 \\ \hline\hline
\multicolumn{7}{l}{$^1$ Average overlap distance} \\
\multicolumn{7}{l}{$^2$ Average overlap volume} \\
\multicolumn{7}{l}{$^3$ There is no relaxation in this scheme} \\
\end{tabular*}%
\end{table}

\begin{figure*}[htbp]
\centering
\includegraphics[width=8.6cm]{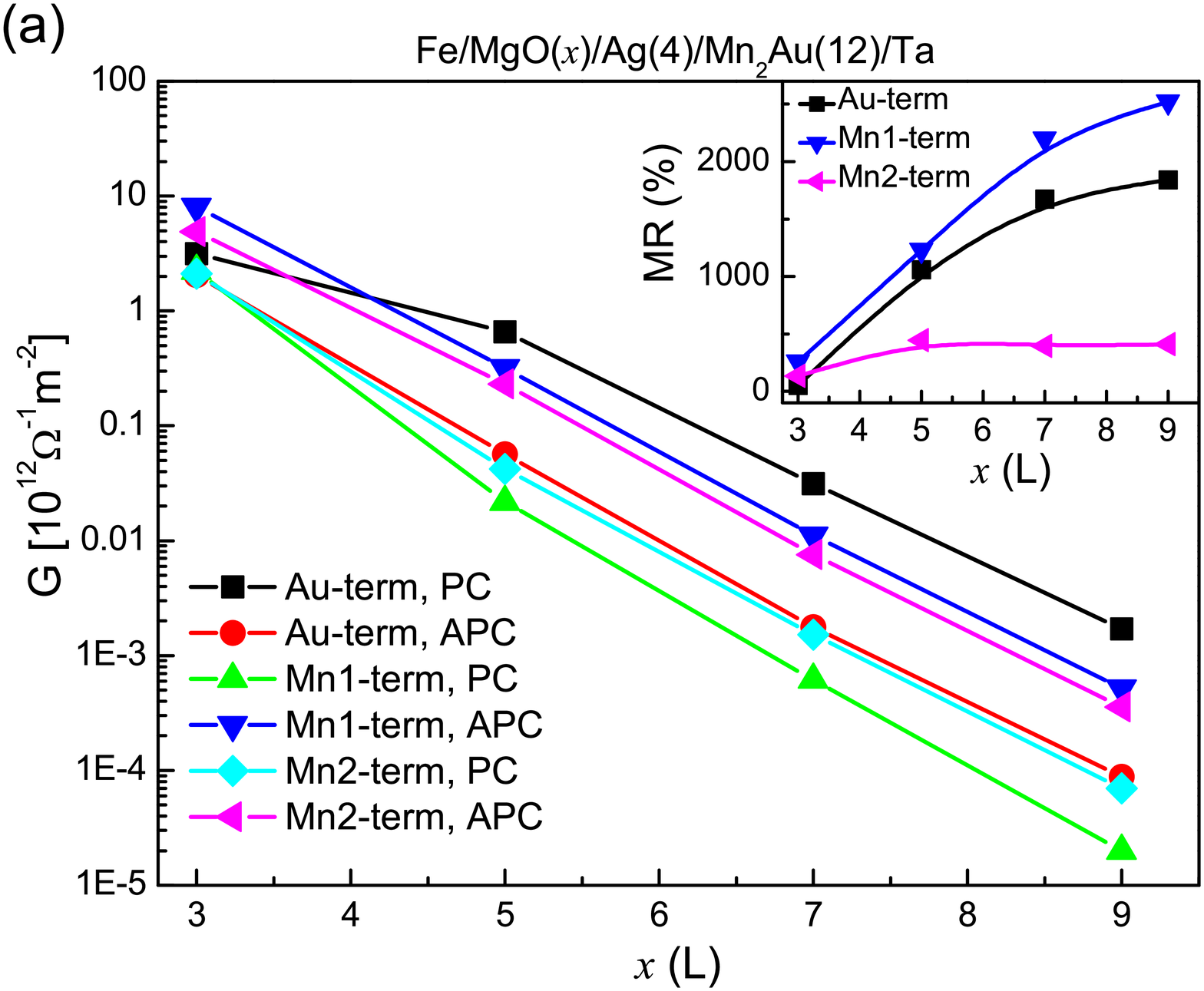}
\includegraphics[width=8.6cm]{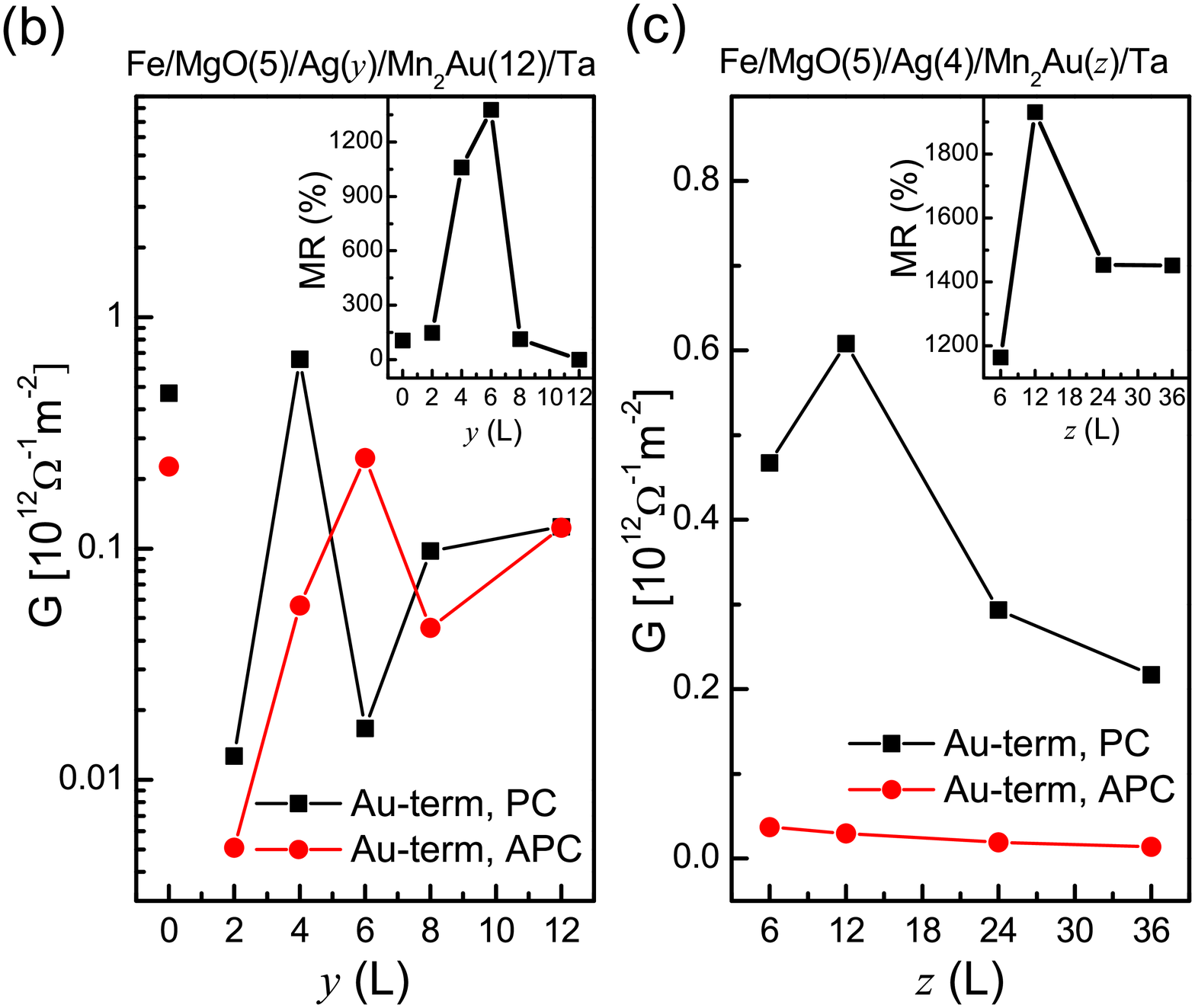}
\caption{Spin transmissions and MRs (inset) in the relaxed Fe/MgO/Ag/Mn$_2$Au/Ta junctions as a functional of (a) MgO barrier, (b) sandwiched Ag, and (c) Mn$_2$Au thickness. For the purpose of comparison, the spin transmissions and MR in the clean Fe/MgO(5)/Mn$_2$Au(12)/Ta junction without relaxation is shown in (b).}
\label{fig2}
\end{figure*}

There are three kinds of Ag/Mn$_2$Au interfaces with regards to the termination atoms, and this can be labeled as Au-term, Mn1-term, and Mn2-term structures with Au, Mn with a positive magnetic moment, and Mn with negative magnetic moment contacted with Ag, respectively. The tetragonal Mn$_2$Au shows easy-plane anisotropy with a stable N{\'e}el vector along the $<110>$ direction.\cite{shick} Thus, the Fe/MgO/Ag/Mn$_2$Au junction would be magnetically stable with a relative angle of $\theta=\pi/4$ between the magnetization of Fe and N{\'e}el order of Mn$_2$Au. Here, the MR is defined as $MR=max[R(0),R(\pi)]/min[R(0),R(\pi)]-1$ with resistance $R=1/G$ and conductance $G=(\unit{e}^2/\unit{h})Tr({tt^{\dag}})$, where $t$ is the transmission part of the scatter matrix $S$.

The spin transport in the nano-structures is sensitive to the interfaces. When the large lateral supercell is used in the Fe/MgO/Ag/Mn$_2$Au/Ta junction, it is hard to relax it by first-principle quantum mechanical calculations. Here, we take several relaxation schemes to find the effect of relaxation on the spin transports of the Mn$_2$Au-based junction, as shown in Table \ref{tab2}. Here, the magnetization of left-side Fe parallels(antiparallels) to the N{\'e}el vector of the right-side Mn$_2$Au as PC(APC) structures. The conductance of both the PC and APC structures of the Fe/MgO(5)/Ag(4)/Mn$_2$Au(12)/Ta junction does not show a simple relation with the minimal distance between the Ag atoms and the Au(Mn) atoms in the Mn$_2$Au layer, and the MRs of the junction are within the range of $950\%$ to $1310\%$ with difference around $27\%$, where the numbers in the bracket indicate the thickness in atomic layers (Ls). This number can be considered as the error bar of the MRs in the calculations. The average overlap distance is related to the interfacial tension, which increases as the minimal AgAu(Mn) distance increases. The average overlap volume is related to the interfacial potential energy, and a minimum was obtained as the minimal distance between Ag atoms and Au(Mn) atoms was around $2.08~\AA$. In the following, we consider this case (the scheme III in Table \ref{tab2}) as the relaxed structure.

Figure \ref{fig2} shows the transmissions and MRs of the relaxed Fe/MgO/Ag/Mn$_2$Au/Ta junctions with respect to the thicknesses of the MgO barrier, the sandwiched Ag, and the Mn$_2$Au layers. The conductance of these AF-MTJs decrease exponentially as the thickness of MgO barrier $x$ increases with a small deviation at $x=3$, and the MRs of these junctions with Au-term and Mn1-term structures increase as $x$ increases while that of the Mn2-term structure achieves is quickly saturated, as shown in the Fig. \ref{fig2} (a) and its inset. The largest MR $\sim2500\%$ was observed in the relaxed Fe/MgO(9)/Ag(4)/Mn$_2$Au(12)/Ta junction in the Mn1-term structure, and it can be compared to that in the ideal MgO-based F-MTJs with same barrier thickness.\cite{butler2001,mathon2001theory} The angular dependence of the MRs in these AF-MTJs was also studied, and a trigonometric function was observed for the junctions with a thicker barrier, and a small deviation was observed in the junctions with the thinner barrier such as $x=3$.

Figure \ref{fig2} (b) shows the sandwiched Ag thickness-dependent transmissions and MRs of the relaxed Fe/MgO(5)/Ag/Mn$_2$Au(12)/Ta junction with Au-term structure. The transmissions of both the PC and APC states of the AF-MTJs are sensitive to the thickness of sandwiched Ag layer $y$, and the largest transmission was at $y=4$ for the PC state and at $y=6$ for the APC state, respectively. The largest MR was at $y=4$ while almost zero MR was recorded at $y=12$. The sandwiched Ag thickness-dependent transmissions of the AF-MTJs with Mn1-term and Mn2-term structures followed the same trend than that of the Au-term structure. The transmission sensitivity of the AF-MTJs with respect to the sandwiched Ag thickness indicates the presence of QW resonance, which could be responsible for the unusual large MR in the relaxed Fe/MgO(5)/Ag$(4)$/Mn$_2$Au(12)/Ta junctions.

Figure \ref{fig2} (c) shows the Mn$_2$Au thickness-dependent transmissions and MRs of the relaxed Fe/MgO(5)/Ag(4)/Mn$_2$Au/Ta junction with the Au-term structure. The peak transmission in the PC state appears at the Mn$_2$Au thickness of $z=12~$L, and the transmission in the APC state decreases as $z$ increases. By and large, the MRs of these junctions are insensitive to the Mn$_2$Au thickness.

\begin{figure*}[tbp]
\centering
\includegraphics[width=18cm]{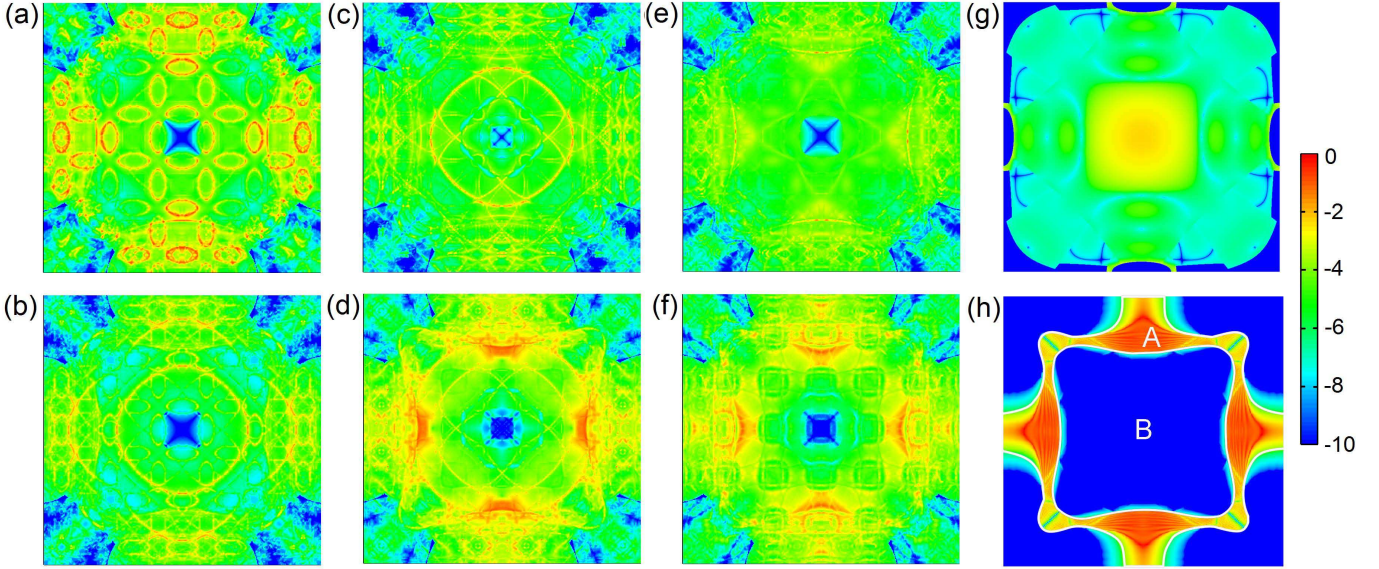}
\caption{$k_{||}-$resolved transmission of the relaxed clean Fe/MgO(5)/Ag(4)/Mn$_2$Au(12)/Ta junction. The upper panel is the PC junction with (a) Au-term, (c) Mn1-term, and (e) Mn2-term structures, respectively, and the lower panel is the APC junction with (b) Au-term, (d) Mn1-term, and (f) Mn2-term structures, respectively. $k_{||}-$resolved transmission of the clean (g) Fe/MgO(5)/Ag junction and (h) Ag/Mn$_2$Au(36)/Ta are shown for comparison. The shape of region "A" in (h) is close to the projection of the Fermi surface of tetragonal Mn$_2$Au along the $<001>$ direction as shown in Fig. \ref{fig4} (f).}
\label{fig3}
\end{figure*}

To understand the unusual giant MRs in the Mn$_2$Au-based junctions, we unpack the transmission of the lateral supercell structure and map it to a $1\times1$ cell. Figure \ref{fig3} gives the $k_{||}-$resolved transmission of the relaxed Fe/MgO(5)/Ag(4)/Mn$_2$Au(12)/Ta junction, and that of the clean Fe/MgO(5)/Ag and Ag/Mn$_2$Au(36)/Ta junctions are also shown for comparison. Therein, the hot spots in the 2d BZ dominate the transmission of the Fe/MgO(5)/Ag(4)/Mn$_2$Au(12)/Ta junctions, which are several orders of magnitude larger than that of the clean Fe/MgO(5)/Ag junction at the same positions. For the Au-term structure, the spin-dependent resonant tunneling (SDRT) occurring in the PC junction is stronger than that in the APC junction, while the SDRT in the APC junction is stronger than that in the PC junction for both the Mn1-term and Mn2-term structures. The distributions and the intensities of the hot spots in the 2d BZ are completely different in the three termination structures, indicating that the termination atoms dominated the SDRT in the junctions.

\begin{figure*}[tbp]
\centering
\includegraphics[width=8.6cm]{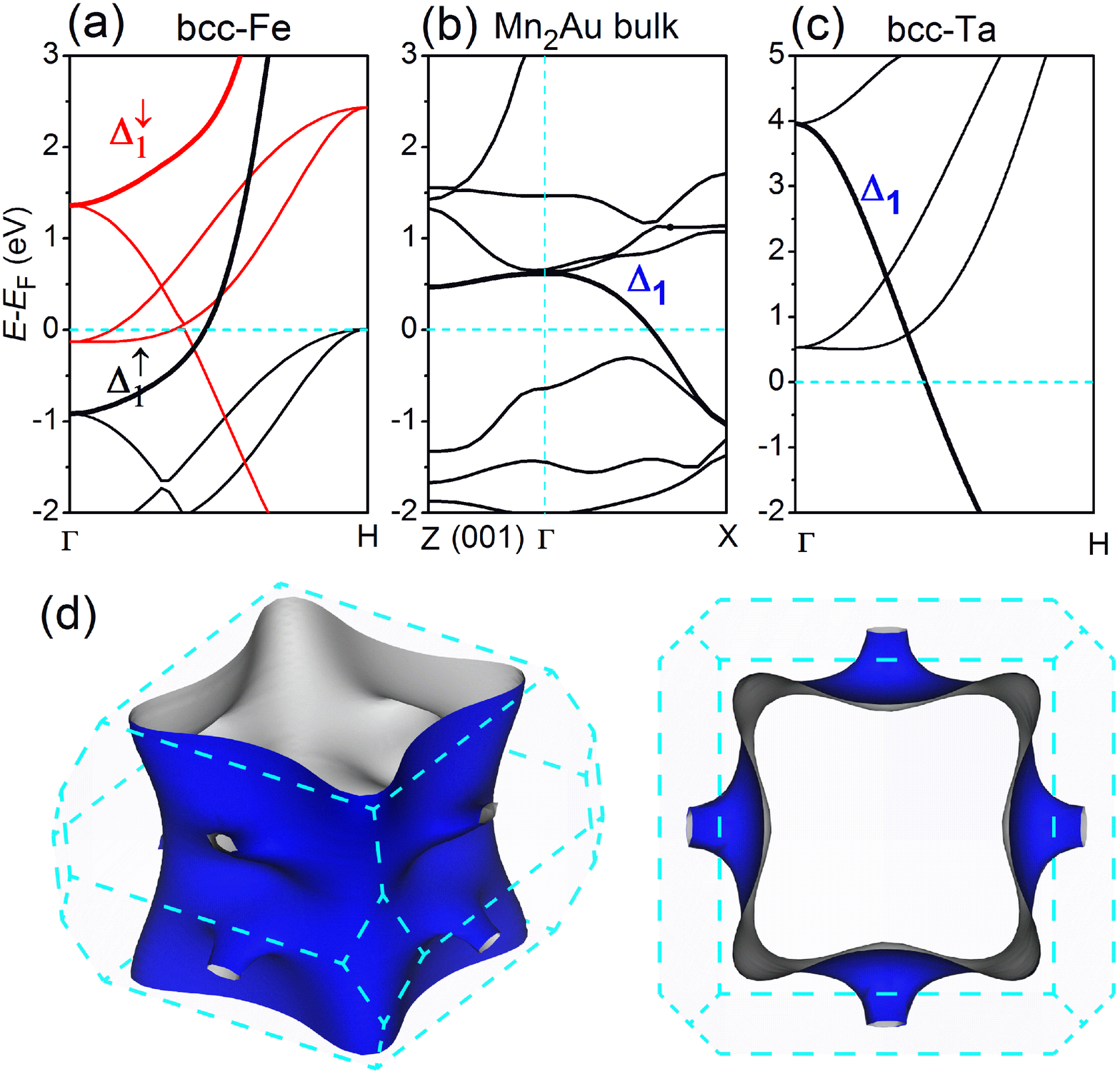}
\includegraphics[width=7.7cm]{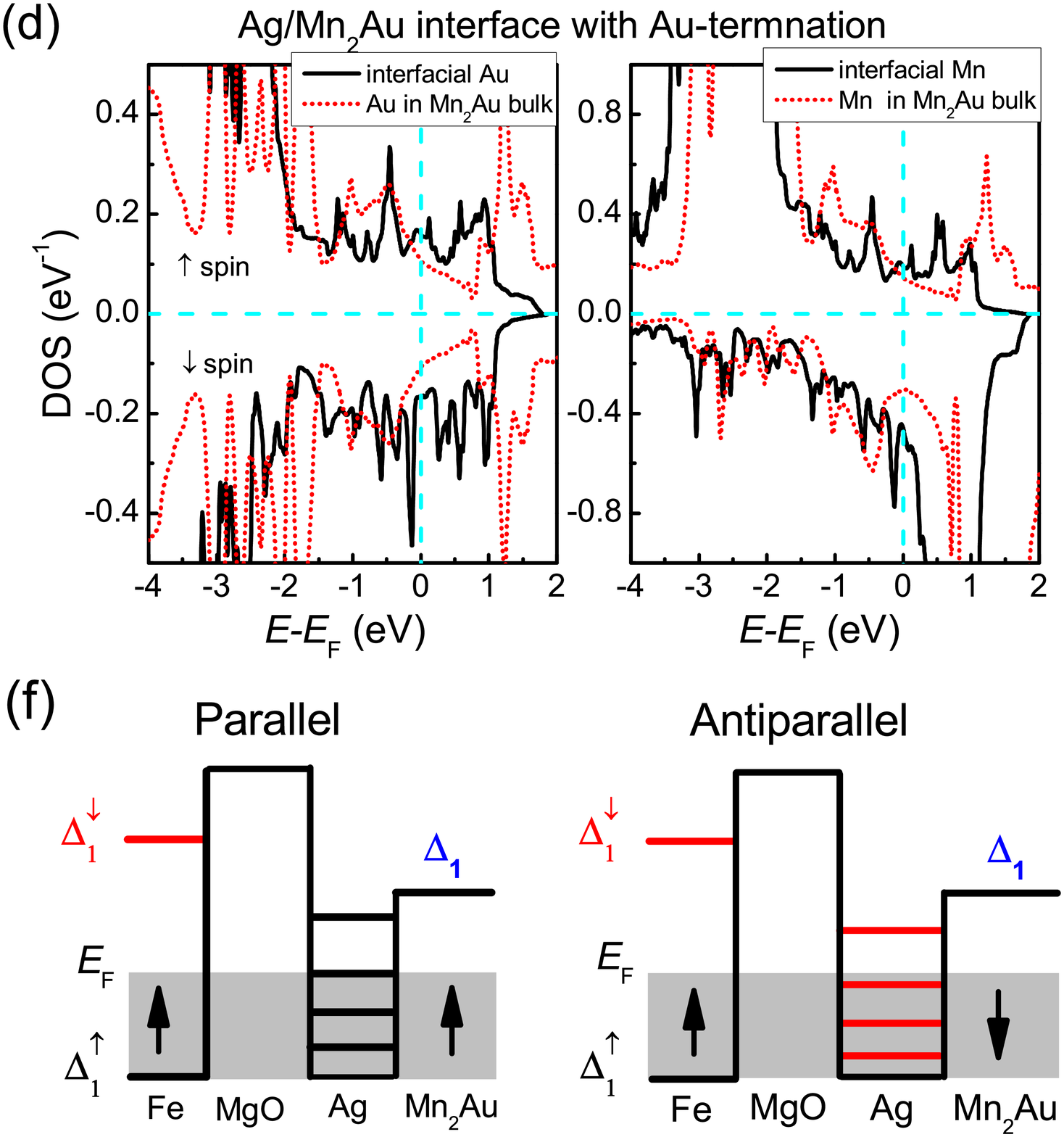}
\caption{Band structures of (a) bcc-Fe, (b) tetragonal Mn$_2$Au, and (c) bcc-Ta with the Fermi level fixed at 0 eV (indicated by the horizontal dotted line). (d) Local density of states (DOSs) of the interfacial Au and Mn1 atoms in the Ag/Mn$_2$Au interface with Au-term structure. (e) 3D view and top view of the Fermi surface of the tetragonal Mn$_2$Au. (f) The potential profiles and schematic illustrations of Fe/MgO(5)/Ag(4)/Mn$_2$Au(12)/Ta(001) junction with Au-term structure. The left and right thicker arrows represent the magnetization of Fe and N{\'e}el order of Mn$_2$Au, respectively. The thicker red and black lines represent the potentials for spin-up and spin-down states of $\Delta_1$ symmetry, respectively.}
\label{fig4}
\end{figure*}

\begin{figure}[t]
\centering
\includegraphics[width=8.6cm]{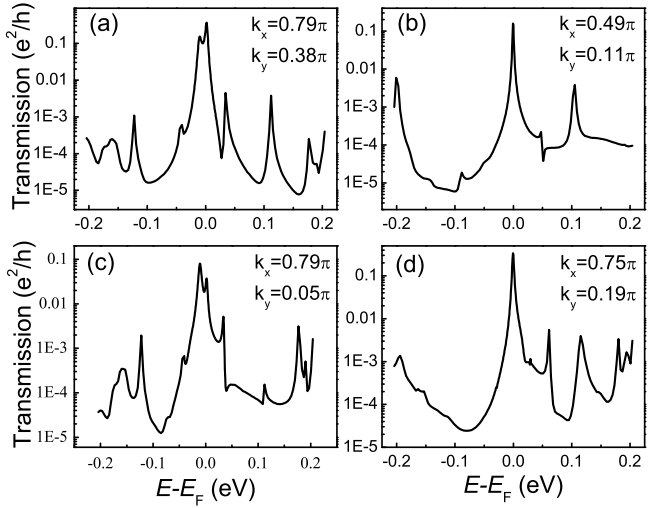}
\caption{Energy dependency of the transmissions of four typical $k_{||}$ points in the 2d BZ of the relaxed Fe/MgO(5)/Ag(4)/Mn$_2$Au(12)/Ta(001) junction with Au-term structure.}
\label{fig5}
\end{figure}

As shown above, it is the differences of the numbers and the intensity of the hot spots in the 2d BZ between the PC and APC states that determine the MRs in these Mn$_2$Au-based junctions. To understand the underlying mechanism, we give the band structures of bcc-Fe, tetragonal Mn$_2$Au, and bcc-Ta, as shown in Fig. \ref{fig4} (a)-(c). For the Mn$_2$Au, only $\Delta_1$ band crosses the Fermi level, and a gap was observed from $\Gamma$ to $Z$ point with a barrier height of $\sim0.65~\unit{eV}$ at the former and $\sim0.48~\unit{eV}$ at the latter, respectively. The Fermi surface of Mn$_2$Au was in the form of a distorted squared bottomless bucket with eight humps around the $N$ point, as shown in Fig. \ref{fig4} (e), indicating that the $s$ $p$, and $d$ states at the Fermi level are highly hybridized. Consequently, a right-going electron across the sandwiched Ag layer would experience a $k_{||}-$dependent barrier. Owing to the symmetry filtering effect of the MgO barrier,\cite{butler2001} only the $\Delta_1$ electrons in the majority ($\uparrow$) spin channel can tunnel through the MgO barrier, then it would be scattered by the $k_{||}-$dependent Mn$_2$Au barrier. This $k_{||}-$dependent QW structure could be responsible for the hottest spots in the 2d BZ of the relaxed Fe/MgO(5)/Ag(4)/Mn$_2$Au(12)/Ta junctions, as shown in the Fig. \ref{fig3} (a)-(f). The QW states would enhance the SDRT in both the "A" and "B" regions of the 2d BZ in these junctions, as shown in Fig. \ref{fig5} also, where the "A" region is the projection of the Fermi surface of tetragonal Mn$_2$Au along the $<001>$ direction, and "B" region is the other of the 2d BZ, as shown in \ref{fig3} (h). For the junctions with thicker sandwiched Ag and thosewithout sandwiched Ag, where the QW structure was ineffective, the SDRT via the interfacial resonant state dominates the transmission.

The spin filtering effect at the Fe/MgO interface could be responsible for the large current spin polarization in both the PC and APC states of these relaxed Mn$_2$Au-based junctions. While the spin-dependent interfacial resonant states (IRSs)\cite{tiusan2004} could be responsible for the giant MRs in the relaxed Fe/MgO/Ag(4)/Mn$_2$Au(12)/Ta junctions as shown in the Fig. \ref{fig4} (d). The local DOSs of both the interfacial Au and Mn atoms of the relaxed Fe/MgO/Ag/Mn$_2$Au/Ta junctions with Au-term structure show sharp peaks around the Fermi level, which may be identified to be the IRSs. Due to the strong orbital hybridization between the interfacial Au and Mn atoms, the IRSs of the former are spin-dependent. These spin-dependent IRSs would couple to the $\Delta_1$ states of the left-side Fe and right-side Ta, leading to SDRT, which could account for the MRs in the Mn$_2$Au-based AF-MTJs.

For the junctions with thinner sandwiched Ag such as the four layers where the QW states are created, these QW states will spin-split by coupling with the IRSs, as shown in Fig. \ref{fig4} (f). When the QW states align with the Fermi level, it would couple to the $\Delta_1$ states of both the left-side Fe and right-side Ta efficiently, resulting in enhanced transmission. The spin splitting of these QW states would be small as a result of the weak coupling. As a result, the voltage bias dependent MRs of the Mn$_2$Au-based AF-MTJs would be complicated, especially when the voltage bias is large. For the relaxed Fe/MgO(5)/Ag(4)/Mn$_2$Au(12)/Ta(001) junction, the QW states favor the PC state than the APC state of the junction with Au-term structure at the equilibrium state. While the QW states favor the APC state than the PC state of the junctions with Mn1 and Mn2-term structures, as shown in the Fig. \ref{fig3}. For the QW states that are sensitive to the thickness of sandwiched metal, the change of the thickness of the sandwiched Ag would affect the energy levels of the QW structure, leading to a  complicated Ag thickness-dependent transmissions and MRs in the Fe/MgO(5)/Ag/Mn$_2$Au(12)/Ta(001) junction, as shown in Fig. \ref{fig2} (b).

Figure \ref{fig5} gives the energy-dependent transmissions of four typical $k_{||}$ points in the 2d BZ of the relaxed Fe/MgO(5)/Ag(4)/Mn$_2$Au(12)/Ta(001) junction with Au-term structure. The $k_{||}$ points in Fig. \ref{fig5} (a) and (b) are within the "B" region of Fig. \ref{fig3} (h), and the other two $k_{||}$ points are within the "A" region. Among them, the $k_{||}$ points in Fig. \ref{fig5} (a) and (c) show two closed peaks around the Fermi level, which corresponds to the bonding and anti-bonding hybrids between the interface states on both sides of the MgO barrier.\cite{PhysRevB.65.064425} While the $k_{||}$ points in the Fig. \ref{fig5} (b) and (d) demonstrate the character of the QW resonant tunneling. For the junctions with thicker (or zero) sandwiched Ag, the QW state is ineffective, and the SDRT via the IRSs contributes mainly to the MRs. From the sandwiched Ag dependent transmissions and the MRs,  we can see that the SDRT via the QW states contributes more to MRs in the relaxed Fe/MgO(5)/Ag(4)/Mn$_2$Au/Ta(001) junction than that via the IRSs as shown in the Fig. \ref{fig2} (b). For the case with thicker Mn$_2$Au, such as 32 monolayers, both the IRSs and the QW states contribute to the SDRT of the junctions, and the former seems to contribute more than the latter. For the case with thinner Mn$_2$Au such as six monolayers, the small barrier height combined with thinner barrier thickness make most $k_{||}$ points of the Fe/MgO(5)/Ag(4)/Mn$_2$Au/Ta junction incapable of QW resonant tunneling, and the IRSs contribute more to the SDRT of this junction. This could be the basic mechanism of the strange thickness dependence of the Mn$_2$Au layer to the relaxed Fe/MgO(5)/Ag(4)/Mn$_2$Au/Ta junctions as shown in Fig. \ref{fig2} (d).

Imperfections considerably affect the spin transports in the Mn$_2$Au-based AF-MTJs. Two kinds of imperfections are assessed here, one is the interfacial OVs and another interfacial AgAu(AgMn) alloys. For the relaxed Fe/MgO(5)/Ag(4)/Mn$_2$Au(12)/Ta junction with Au-term structure, $10\%~$OVs at the MgO/Ag interface can reduce the MR from $1150\%$ in the clean junction to $22\%$ in the dirty junction, and 2 L of a randomly disordered $50\%-50\%$ interfacial alloys exited at the Ag/Mn$_2$Au interface can reduce the MR to around $400\%$. That is to say, the AF-MTJs are more sensitive to interfacial OVs than to the interfacial AgAu(AgMn) alloys. So, to achieve larger MRs in the Mn$_2$Au-based tunnel junction, the interfaces should be as clean as possible. Furthermore, the thermal effect on the spin transports at room temperature was also studied. MR reduction of $\sim20\%$ was observed in the relaxed Fe/MgO(5)/Ag(4)/Mn$_2$Au(12)/Ta junction with Au-term structure, where the QW resonant tunneling should account for this insensitivity to the temperature-induced lattice disorder.

Although the MR in the AFM-based spin valves has been theoretically predicted for a long time,\cite{nunez2006theory,xu2008spin,saidaoui2014spin,jia2017structure,PhysRevApplied.12.044036} there is no experimental evidence up till date. The poor interfaces could be responsible for the large difference between the theoretical and experimental observations. The existence of lattice mismatch between the antiferromagnetic metals such as IrMn and FeMn alloys and the sandwiched nonmagnet such as Cu would unavoidably introduce imperfections such as interfacial disorder and impurities. These imperfections would destroy the interfacial states, and then destroy the MR effect in these spin valves. A good buffer layer is helpful to retrieve the interfacial states, and then restore the MR in the AFM-based spin valve.

\section{Summary}

Summarily, we calculated the spin transports in the Fe/MgO/Ag/Mn$_2$Au/Ta junctions based on first-principle scattering theory. Giant MR more than $1000\%$ was observed in some junctions, which is comparable to that in the MgO-based F-MTJs. The unusual giant MRs in the Mn$_2$Au-based AF-MTJs could be related to 1) the spin filtering effect of the Fe/MgO interface, 2) band structure of tetragonal Mn$_2$Au, and 3) spin-dependent IRSs at the Ag/Mn$_2$Au interface. The $k_{||}$ dependent energy barrier of tetragonal Mn$_2$Au is the basis of QW resonant tunneling. The interplay of the QW states in the sandwiched Ag and the IRSs of the Ag/Mn$_2$Au interface make the QW resonant tunneling spin-dependent, and this accounts for the giant and robust MR in the junctions with the thinner sandwiched Ag. The MRs in the Mn$_2$Au-based AF-MTJs are sensitive to the interfacial OVs, while they are not so sensitive to the interfacial AgAu(AgMn) alloys and temperature-induced lattice disorders.

\section{ Acknowledgment}

We gratefully acknowledge the financial support from the National Natural
Science Foundation of China (Grant No. 11804062, and 11804082), and the National Key Research and Development Program (Grant no. 2018YFB0407600).

\end{document}